\documentclass[12pt]{article}
\usepackage{amsfonts}
\usepackage{amsmath}
\usepackage{graphicx}
\usepackage{epstopdf}
\newcommand{\be}{\begin{equation}}
\newcommand{\ba}{\begin{eqnarray}}
\newcommand{\ee}{\end{equation}}
\newcommand{\ea}{\end{eqnarray}}
\newcommand{\cosech} { {\rm cosech}}

\begin{document}

\title{Solutions of (1+1)-dimensional Dirac equation associated with exceptional orthogonal polynomials and the parametric symmetry}

\author{Suman Banerjee$^{a}$\footnote{e-mail address: suman.raghunathpur@gmail.com(S.B)}, Rajesh Kumar Yadav$^{a}$\footnote{e-mail address: rajeshastrophysics@gmail.com(R.K.Y)}, 
Avinash Khare$^{b}$\footnote {e-mail address: avinashkhare45@gmail.com (A.K)}, 
Nisha Kumari$^{c}$\footnote{e-mail address: nishaism0086@gmail.com (N.K)} and \\
  Bhabani Prasad Mandal$^{d}$\footnote{e-mail address: bhabani.mandal@gmail.com (B.P.M).}}
 \maketitle
{$~^a$Department of Physics, S. K. M. University, Dumka-814110, India.\\
$~^b$Department of Physics, Savitribai Phule Pune University, Pune-411007, India.\\
$~^c$Department of Physics, S. P. College, Dumka-814101, India.\\
$~^d$Department of Physics, Banaras Hindu University, Varanasi-221005, India. }

\begin{abstract}
We consider $1+1$-dimensional Dirac equation with rationally extended scalar
potentials corresponding to the radial oscillator, the trigonometric Scarf 
and the hyperbolic Poschl-Teller potentials and obtain their solution in terms
of exceptional orthogonal polynomials. Further, in the case of the 
trigonometric Scarf and the hyperbolic Poschl-Teller cases, new family of 
Dirac scalar potentials are generated using the idea of parametric symmetry 
and their solutions are obtained in terms of conventional as well as 
exceptional orthogonal polynomials.

\end{abstract}

\section{Introduction}

Dirac equation plays an important role 
in the study of the dynamics of the relativistic systems with spin $\hbar/2$.
Dirac equation has been applied to solve 
many problems in nuclear and high energy physics \cite{nuc,gino_1,gino_2}. 
In the quantum mechanical context, by now solutions of the Dirac equation 
have been obtained in the case of 
several scalar and vector potentials \cite{fra,gbjw,dfl,era,nki3,
pct,gth} using different approaches such as 
supersymmetric quantum mechanics (SQM) approach, Nikiforov-Uvarov approach, the
point canonical transformation approach, the group theoretic approach etc. 
Dirac equation has been solved for a broad class of potentials 
such as the Morse potential \cite{mrse1}, the Coulomb potential \cite{clm}, the
P\"oschl-Teller potential \cite{pst}, the Hulthen potential \cite{hul} and the 
Scarf potential \cite{scf} etc. In
the last few decades, it has been observed that the techniques of SQM
\cite{fra,khare,kas,smr,rjv,ah,br,gj} play an important role in solving 
Dirac equation in $1+1$ space time in the case of various scalar potentials.

In recent years, the discovery of two new orthogonal polynomials namely the $X_{m}$ exceptional Laguerre
and $X_{m}$ exceptional Jacobi orthogonal polynomials \cite{dnr1,xm1,xm2} (where the degree $m\geq 1$ instead of zero 
as in the case of usual polynomials) has lead to the discovery of rational 
extensions of several exactly solvable potentials in non-relativistic QM. In 
particular, the solution of the Schr\"odinger equation corresponding to the 
rationally extended
potentials have been obtained in terms of exceptional orthogonal polynomials 
(EOPs) \cite{que}-\cite{rks} or in the form of combination of usual orthogonal 
polynomials \cite{op3,yg15}. 

Another development in non-relativistic QM is that of the parametric symmetry \cite{para2,para1} .
It has been shown that 
while for some potentials the parametric symmetry leads to another set of
solutions keeping the same form of the conventional potentials, 
but in the case of the corresponding rationally extended potentials
this symmetry generates  another form of the extended potentials and hence
completely different solutions. 

In contrast, in the relativistic case only few attempts 
have been made so far to solve the Dirac equation corresponding to the 
rationally extended scalar potentials \cite{asb,as,kh}. Besides, the role of
the parametric symmetry has not been explored in the Dirac case. The 
purpose of this paper is to obtain solutions of the Dirac equation 
in the case of few rationally extended scalar potentials and also study the
role of the parametric symmetry in Dirac equation with scalar potentials.

In particular, in this paper, we consider the $(1+1)$-dimensional Dirac 
equation with three different forms of the scalar potentials $\tilde{\phi}(x)$ 
whose solutions are well known. We consider the corresponding 
rationally extended Dirac scalar potentials and obtain their solution in the 
form of the EOPs. Further, we show that
similar to the Schr\"odinger case, there is also a parametric symmetry in the
case of some of the Dirac equation with some of the scalar potentials.
In particular, extending the idea of the parametric symmetry discussed in 
\cite{para2,para1} to the relativistic 
case, we generate a family of new form of the conventional as well as the 
rationally extended scalar potentials and obtain the solution of the 
corresponding Dirac equation.

The plan of the paper is as follows. In Section $2$, we review the $1+1$ 
dimensional Dirac equation with scalar potential and discuss how its solutions
can be obtained using SQM approach. In Sec. $3$ we obtain the solution of the 
Dirac equation with rationally extended radial oscillator, trigonometric Scarf 
potential and the hyperbolic P\"oschl-Teller potential and obtain their 
solutions in terms of EOPs using the 
SQM approach. In all these cases, for simplicity we first obtain solutions 
in terms of the $X_{1}$ Jacobi or $X_1$ Laguerre polynomials and then 
generalize to the general the $X_{m}$ case. In Sec. $4$, we show that the 
trigonometric Scarf and the hyperbolic Poschl-Teller Dirac problems have novel 
parametric symmetry. Using this symmetry, we obtain another form 
of the rationally extended Dirac scalar potentials and obtain their 
solutions. Finally, in Sec. $5$ we summarize our results and point out 
few open problems.
   
\section{Formalism}

In this section, we review the solutions of the Dirac equation with general 
scalar potentials in $1+1$ dimension and show how the problem can be reduced to 
two decoupled Schr\"odinger equations. In this way, using the well known SQM 
approach one can obtain the exact solutions of the corresponding Dirac 
problems in several cases.

The Dirac Lagrangian in $1+1D$ with a 
Lorentz scalar potential $\tilde{\phi}(x)$ is given by
\be\label{vfc}
L =i \bar{\Psi}\gamma^{\mu }\partial _{\mu }\Psi 
-\tilde{\phi} (x)\bar{\Psi }\Psi, \qquad \mu=0,1 \,  
\ee 
where $\Psi$ is the Dirac spinor. The Dirac equation following from 
Eq. (\ref{vfc})
is
\be\label{sda} 
 i \gamma^{\mu }\partial _{\mu }\Psi(x,t) - \tilde{\phi}(x) \Psi(x,t) =0\,.
\ee
Let
\be\label{energy}
\Psi (x,t) =\exp(-i\varepsilon  t)\xi (x)\,,
\ee
so that the above Dirac equation reduces to
\be\label{pyt}
\gamma^{0}\varepsilon\xi (x)+i\gamma ^{1}\frac{d}{dx}\xi(x)
-\tilde{\phi}(x)\xi (x)=0\,.
\ee
Now, we choose the following $2D$ representation of the gamma matrices i.e, 
\be\label{mno}
\gamma ^{0}=\sigma_{x}=
\begin{bmatrix}
0&1\\
1&0\\
\end{bmatrix},
\qquad 
\gamma ^{1}=i\sigma_{z}=
\begin{bmatrix}
i&0\\
0&-i\\
\end{bmatrix}
\mbox{and} \qquad 
\xi (x)=
\begin{bmatrix}
\tilde{\Psi} ^{(1)}(x)\\
\tilde{\Psi} ^{(2)}(x)\\
\end{bmatrix}, 
\ee
 and get two coupled equations
\be
\frac{d}{dx}\tilde{\Psi} ^{(1)}(x)+\tilde{\phi}(x)\tilde{\Psi}^{(1)}(x)
= \varepsilon \tilde{\Psi} ^{(2)}(x)\,,
\ee
and
\be
\frac{d}{dx}\tilde{\Psi} ^{(2)}(x)-\tilde{\phi}(x)\tilde{\Psi}^{(2)}(x) 
= -\varepsilon \tilde{\Psi} ^{(1)}(x)\,.
\ee
These two equations can be decoupled easily and we obtain
\be\label{ghf}
-\frac{d^{2}}{dx^{2}}\tilde{\Psi} ^{(1)}(x)+\tilde{V}^{(1)}(x)\tilde{\Psi} ^{(1)}(x)=\varepsilon ^{2}\tilde{\Psi}^{(1)}(x)
\ee
and
\be\label{ght}
-\frac{d^{2}}{dx^{2}}\tilde{\Psi} ^{(2)}(x)+\tilde{V}^{(2)}(x)\tilde{\Psi} ^{(2)}(x)=\varepsilon ^{2}\tilde{\Psi} ^{(2)}(x)
\ee
respectively. These two equations (\ref{ghf}) and (\ref{ght}) are equivalent to
two independent Schr\"odinger equations with potentials 
 
\be\label{wsa}
\tilde{V}^{(1,2)}(x) = \tilde{\phi}^{2}(x)\mp \tilde{\phi}^{'}(x).
\ee
The solutions of these equations can be easily obtained for several 
$\tilde{\phi} (x)$ using the well known SQM approach \cite{khare} by defining 
two operators $\hat{A}$ and $\hat{A^{\dagger }}$ as
  
\be\label{pou}
\hat{A} = \frac{d}{dx}+\tilde{\phi}(x) \quad \mbox{and} \quad \hat{A^{\dagger }} = -\frac{d}{dx}+\tilde{\phi}(x).
\ee
In this way, the Eqs. (\ref{ghf}) and (\ref{ght}) are reduced to 
\be\label{tyf}
\hat{A^{\dagger}}\hat{A}\tilde{\Psi} ^{(1)} = \varepsilon ^{2}\tilde{\Psi} ^{(1)}\quad \mbox{and} \quad \hat{A}\hat{A}^{\dagger}\tilde{\Psi} ^{(2)} = \varepsilon ^{2}\tilde{\Psi} ^{(2)}
\ee
respectively. On comparing with the well known formalism of SQM \cite{khare}, 
we see that there is a supersymmetry in the problem and the scalar potential 
$\tilde{\phi}(x)$ is just the superpotential of the Schr\"odinger formalism.
Further $\tilde{\Psi}^{(1)}$ and $\tilde{\Psi}^{(2)}$ 
are the eigen functions of the 
Hamiltonians $H_{1}\equiv \hat{A^{\dagger}}\hat{A}$ and 
$H_{2}\equiv \hat{A}\hat{A^{\dagger}}$ respectively with $\tilde{V}^{(1,2)}(x)$
being the partner potentials. Thus the eigenvalues and the eigen functions of
the two Hamiltonians are related except that one of them has an extra
bound state at zero energy so long as $\tilde{\phi}(x \rightarrow \pm \infty)$ 
have opposite signs. Without loss of generality we shall always choose 
$\tilde{\phi}(x)$ such that the ground state energy of $H_1$ is zero.
In that case the eigen functions and eigenvalues ($\tilde{E}^{(1)}_{n}$ and 
$\tilde{E}^{(2)}_{n}$)  corresponding to these two Hamiltonians are related
to each other as follows \cite{khare}
\be\label{lkh}
\tilde{\Psi}^{(2)}_{n}(x)
=[\tilde{E}^{(1)}_{n+1}]^{-\frac{1}{2}}\hat{A}\tilde{\Psi}^{(1)}_{n+1}(x)\,,
\ee

\be\label{juhg}
\tilde{\Psi}^{(1)}_{n+1}(x)
=[\tilde{E}^{(2)}_{n}]^{-\frac{1}{2}}\hat{A}^{\dagger }\tilde{\Psi}^{(2)}_{n}(x)
\ee
and
\be\label{bcg}
\tilde{E}^{(2)}_{n}=\tilde{E}^{(1)}_{n+1},\quad \tilde{E}^{(1)}_{0}=0\,.
\ee
Here $n = 0, 1, 2,...$. Thus once we have the eigen functions
$\tilde\Psi^{(1)}_n(x)$ and the energy eigenvalues $\tilde{E}^{(1)}_{n}$, we 
can easily obtain $\tilde\Psi^{(2)}_n(x)$ and $\tilde{E}^{(2)}_{n}$ using 
Eq. (\ref{lkh}) and (\ref{bcg}) respectively. 
 





\section{Rational Dirac Potentials}

We shall now discuss three examples of the rational scalar Dirac potentials
$\tilde{\phi}(x)$, i.e. the radial oscillator, trigonometric Scarf and the
Poschl-Teller potentials. To motivate the discussion we first mention the
well known results about the corresponding conventional Dirac scalar 
potentials and then obtain the solution of the corresponding rational cases. 
For simplicity, we first discuss the $X_1$ case and then 
generalize to the general $X_m$ case.

\subsection{Radial oscillator} 

\subsubsection{The conventional case}

Let us consider the scalar potential defined on the half line 
($0\le r \le \infty$) of the form 
\be\label{uhyt}
\tilde{\phi}(r) \longrightarrow  \tilde{\phi}_{con}(r) 
= \frac{1}{2}\omega r-\frac{\ell+1}{r}\,.
\ee
On using Eq. (\ref{wsa}) it gives rise to 
\ba\label{koji}
\tilde{V}^{(1)}\longrightarrow \tilde{V}^{(1)}_{con}(r)&=& \tilde{\phi}^{2}_{con}(r)-\tilde{\phi}^{'}_{con}(r)\nonumber\\
 &=&\frac{\omega r^{2}}{4}+\frac{\ell(\ell+1)}{r^{2}}-\omega (\ell+\frac{3}{2}),
\ea 
which is the well known radial oscillator potential ( $\omega>0,\ell>0$) whose 
solutions \cite{khare} are given in terms of the classical Laguerre polynomials
$L^{(\ell+\frac{1}{2})}_{n}(z)$ 
\be\label{wfrm}
\tilde{\Psi}^{(1)}_{con,n}(r) = N^{\ell,1}_{con,n}r^{\ell+1}\exp\bigg(-\frac{z(r)}{2}\bigg)L_{n}^{(\ell+\frac{1}{2})}(z(r)), \quad n=0,1,2,...
\ee
where $z(r)=\frac{\omega r^2}{2}$ and the normalization constant  
\be\label{pku}
N^{\ell,1}_{con,n}=\bigg[\frac{n!\omega^{(\ell+\frac{3}{2})}}
{2^{(\ell+\frac{1}{2})}(\ell+n+\frac{1}{2})\Gamma(\ell+n+\frac{1}{2})}\bigg]^{1/2}\,.
\ee 
The corresponding energy eigenvalue are
\be\label{rtd}
\tilde{E}^{(1)}\longrightarrow \tilde{E}^{(1)}_{con,n}
=\varepsilon ^{2}=\omega(2n+\ell+\frac{3}{2})\,.
\ee

\subsubsection{The Rationally Extended Case}

(a) {\bf{The $X_{1}$ Case}} 

In this case, we consider the scalar potential $\tilde{\phi}(r)$ which is 
defined as the sum of the conventional scalar potential 
$(\tilde{\phi}_{con}(r))$ as given by Eq. (\ref{uhyt}) and a rational 
term $\tilde{\phi}_{rat}(r)$ i.e, 
\be\label{bvgf}
\tilde{\phi}(r)\longrightarrow \tilde{\phi}_{ext}(r)
=\tilde{\phi}_{con}(r)+\tilde{\phi}_{rat}(r),
\ee
where 
\be\label{trf}
\tilde{\phi}_{rat}(r)=\frac{4\omega r}{(2z(r)+2\ell+ 1)(2z(r)+2\ell+ 3)}.
\ee 
On using Eq. (\ref{bvgf}) in Eq. (\ref{wsa}), we get the rationally extended 
radial oscillator potential \cite{que}
\be\label{vcdg}
\tilde{V}^{(1)}_{rat}(r)=\tilde{V}^{(1)}_{con}(r)+\tilde{V}^{(1)}_{rat}(r),
\ee
with
\be\label{kvcf}
\tilde{V}^{(1)}_{rat}(r) = 4\omega\bigg(\frac{1}{(2z(r)+2\ell+1)}
-\frac{2(2\ell+1)}{(2z(r)+2\ell+1)^{2}}\bigg)\,,
\ee 
while $\tilde{V}^{(1)}_{con}(r)$ is as given by Eq. (\ref{koji}). 
The solution of the corresponding Schr\"odinger equation is \cite{que}
\be\label{sdfr}
\tilde{\Psi}^{(1)}_{ext,n}(r) = N^{\ell}_{n,ext}\frac{r^{\ell+1}\exp\big(-\frac{z(r)}{2}\big)}{L^{(\ell-\frac{1}{2})}_1(z(r))}\hat{L}_{n+1}^{(\ell+\frac{1}{2})}(z(r)),
\ee
where $\hat{L}^{(\alpha)}_{n+1}(z(r))$
is $X_{1}$ exceptional Laguerre Polynomials while 
the normalization constant is \cite{que}
\be\label{pbtd}
N^{\ell,1}_{ext,n}=\bigg[\frac{n!\omega^{(\ell+\frac{3}{2})}}{2^{(\ell+\frac{1}{2})}(\ell+n+1+\frac{1}{2})\Gamma(\ell+n+\frac{1}{2})}\bigg]^{1/2}.
\ee
Notice that the energy eigenvalues are same as that of the conventional one 
and are given by 
\be\label{rfgh}
\tilde{E}_{n,ext}^{(1)}=\varepsilon ^{2}_{ext}=2n\omega.
\ee

(b) {\bf{The $X_{m}$ Case}}

The above results for the $X_1$ case are immediately generalized to the
general $X_m$ case. In this case the scalar potential is defined as 
$\tilde{\phi}(r)\longrightarrow \tilde{\phi}_{ext,m}(r)$, given by
\be\label{fcxb}
\tilde{\phi}_{ext,m}(r) = \tilde{\phi}_{con}(r) 
+ \tilde{\phi}_{m,rat}(r); \quad (m=0,1,2,....)\,,
\ee
where $\tilde{\phi}_{con}(r)$ is again as given by Eq. (\ref{uhyt}) while 
the rational term $\tilde{\phi}_{m,rat}(r)$ is given by
\be\label{oijk}
\tilde{\phi}_{m,rat}(r)=\omega r\bigg[\frac{L^{(\ell+\frac{1}{2})}_{m-1}(-z(r))}{L^{(\ell-\frac{1}{2})}_{m}(-z(r))}-\frac{L^{(\ell+\frac{3}{2})}_{m-1}(-z(r))}{L^{(\ell+\frac{1}{2})}_{m}(-z(r))}\bigg].
\ee
On using $\tilde{\phi}_{ext,m}(r)$ instead of $\tilde{\phi}_{con}(r)$ in 
Eq. (\ref{wsa}), we get \cite{os,rkyd} 
\be\label{sum1}
\tilde{V}^{(1)}_{ext,m}(r) = \tilde{V}_{con}^{(1)}(r)+\tilde{V}_{rat,m}^{(1)}(r)
\ee
where $\tilde{V}_{con}^{(1)}(r)$ is again as given by Eq. (\ref{koji}) while
\ba\label{jhb}
\tilde{V}^{(1)}_{rat,m}(r)&=&-2\omega z(r)\frac{L^{(\ell+\frac{3}{2})}_{m-2}(-z(r))}{L_{m}^{(\ell-\frac{1}{2})}(-z(r))}+\omega(2 z(r)+2\ell-1)\frac{L^{(\ell+\frac{1}{2})}_{m-1}(-z(r))}{L_{m}^{(\ell-\frac{1}{2})}(-z(r))}\nonumber\\
&+&2\omega z(r)\bigg(\frac{L^{(\ell+\frac{1}{2})}_{m-1}(-z(r))}{L_{m}^{(\ell-\frac{1}{2})}(-z(r))}\bigg)^2 - 2m\omega,\quad 0<r<\infty.
\ea
The solutions of the corresponding Schr\"odinger equation are in terms of 
$X_{m}$ Laguerre Polynomials $(\hat{L}_{n+m}^{(\ell+\frac{1}{2})}(z))$ and 
are given by  
\be\label{wfrm1}
\tilde{\Psi}^{(1)}_{ext,n,m}(r,\omega,\ell) = N^{\ell,1}_{ext,n,m}\frac{r^{\ell+1}\exp\big(-\frac{z(r)}{2}\big)}{L^{(\ell-\frac{1}{2})}_m(-z(r))}\hat{L}_{n+m}^{(\ell+\frac{1}{2})}(z(r)), \quad m=1,2,...,
\ee
where
\be
\hat{L}^{(\alpha)}_{n+m}(z)=L^{(\alpha)}_m(-z)L^{(\alpha-1)}_{n}(z)+L^{(\alpha-1)}_m(-z)L^{(\alpha)}_{n-1}(z); \quad n\geq m
\ee
with the normalization constant being
\be\label{jbhg}
N^{\ell,1}_{ext,n,m}=\bigg[\frac{n!\omega^{(\ell+\frac{3}{2})}}{2^{(\ell+\frac{1}{2})}(\ell+n+m+\frac{1}{2})\Gamma(\ell+n+\frac{1}{2})}\bigg]^{1/2}.
\ee 
As a check on our calculations, for $m = 0$ and $1$, we 
recover the results corresponding to the conventional and the $X_{1}$ extended 
rational scalar Dirac potentials respectively. 
The energy spectrum is again the  same as of the conventional case and given 
by Eq. (\ref{rtd}). 

The plots of the Dirac scalar 
potentials $\tilde{\phi}_{ext,m}(r)$ and   
the corresponding normalized ground state eigen functions 
$\tilde{\Psi}^{(1)}_{ext,0,m}(r,\omega,\ell)$ are given for $m = 0, 1, 2$ in 
figs. $1(a)$ and $1(b)$ respectively in case $\omega=2$, $\ell=1$.

\includegraphics[scale=1.2]{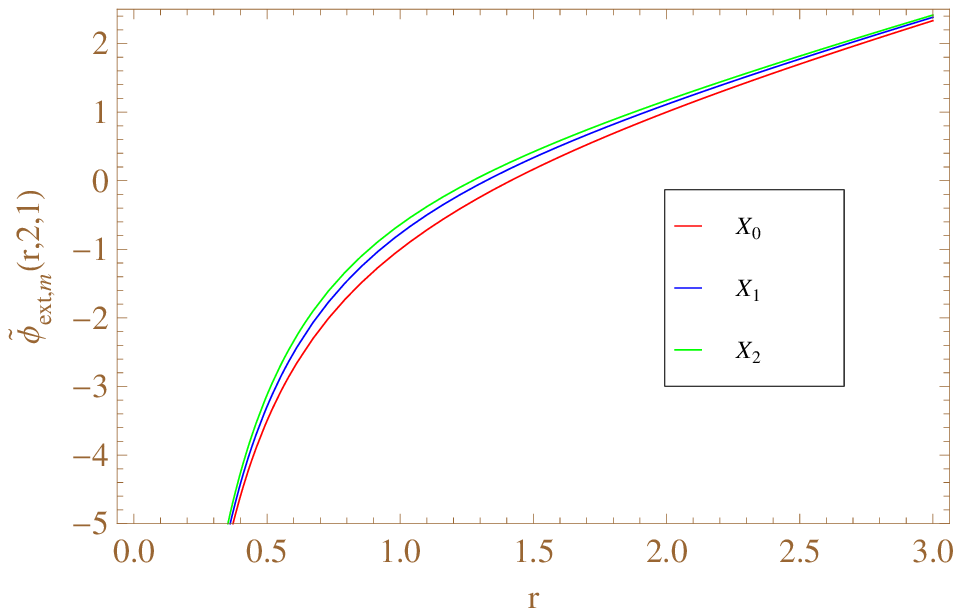}\\ 
{\bf Fig.1}: {(a) {\it Rationally extended Dirac scalar potentials for $m=0,1$ and $2$.}\\\
\includegraphics[scale=1.2]{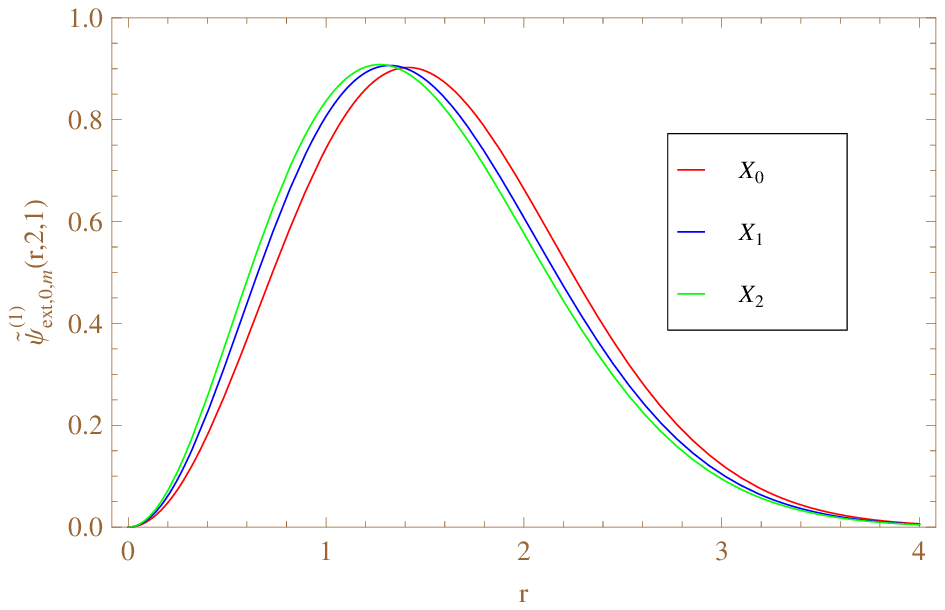} \\
{\bf Fig.1}: {(b) {\it Normalized ground-state wave functions for $m = 0, 1$ and $2$.}\\\ 

\subsection{Trigonometric Scarf case}

\subsubsection{The Conventional Case}

In this case, the scalar potential $\tilde{\phi} (x)\rightarrow 
\tilde{\phi}_{con}(x,A,B)$ (defined on  $-\frac{\pi}{2}\le x \le \frac{\pi}{2}$) is given by
\be\label{dsr}
\tilde{\phi}_{con}(x,A,B) = A\tan{x} -B\sec{x};\quad -\frac{\pi}{2}< x < \frac{\pi}{2}, \quad 0<B<A-1.
\ee
On using this $\tilde{\phi}_{con,A,B}(x)$ in Eq. (\ref{wsa}), we get the well 
known trigonometric Scarf potential \cite{khare}
\be\label{kljh}
\tilde{V}^{(1)}(x)\longrightarrow \tilde{V}^{(1)}_{con}(x,A,B) = [(A-1)A+B^{2}]\sec^{2}{x}-B(2A-1)\sec{x}\tan{x}-A^{2}.
\ee
The solution of the Schr\"odinger equation corresponding to this potential 
is well known and is given in 
terms of the classical Jacobi polynomial $P^{(\alpha,\beta)}_{n}(z)$ as
\be\label{rfdg}
\tilde{\Psi}^{(1)}_{con,n}(x,A,B)=N^{(1)}_{con,n}(\alpha ,\beta)(1-z(x))^{\frac{(A-B)}{2}}(1+z(x))^{\frac{(A+B)}{2}}P^{(\alpha,\beta)}_{n}(z(x)).
\ee
Here $\alpha=A-B-\frac{1}{2}$ , $\beta=A+B-\frac{1}{2}$,  $z(x)=\sin x$ and 
the normalization constant is given by
\be\label{knbgf}
N^{(1)}_{con,n}(\alpha,\beta) = \Bigg[\frac{ n!(\alpha+\beta+2n+1 ) \Gamma (n+\alpha+\beta+1)}{2^{\alpha+\beta+1}(n+\beta)\Gamma (n+\alpha+1)\Gamma (n+\beta)}\Bigg]^{\frac{1}{2}}.
\ee
The energy eigenvalues are
\be\label{vfcg}
\varepsilon ^{2} = \tilde{E}^{(1)}_{con,n}=(A+n)^2 - A^2\,.
\ee

\subsubsection{The Rationally Extended Case}

(a) {\bf{The $X_{1}$ Case:}}  In the extended $X_{1}$ case, the function 
\be\label{ppj}
\tilde{\phi}(x)\longrightarrow \tilde{\phi}_{ext}(x,A,B)=\tilde{\phi}_{con}(x,A,B)+\tilde{\phi}_{rat}(x,A,B)
\ee 
where $\tilde{\phi}_{con}(x,A,B)$ is as given by Eq. (\ref{dsr}) while
\be\label{jjp}
\tilde{\phi}_{rat}(x,A,B) = -2Bz^{'}(x)\bigg[\frac{1}{2A-1-2Bz(x)}-\frac{1}{2A+1-2Bz(x)}\bigg].
\ee
Here $z'(x)$ is the first derivative of $z(x)$ with respect to $x$. 
Using Eq. (\ref{wsa}), the corresponding rationally extended trigonometric 
Scarf potential $\tilde{V}^{(1)}_{ext}(x,A,B)$ turns out to be \cite{que} 
\be\label{pjbk}
\tilde{V}^{(1)}_{ext}(x,A,B) = \tilde{V}^{(1)}_{con}(x,A,B)+\tilde{V}^{(1)}_{rat}(x,A,B)
\ee
where $\tilde{V}^{(1)}_{con}(x,A,B)$ is as given by Eq. (\ref{kljh}) while
\be\label{uijk}
\tilde{V}^{(1)}_{rat}(x,A,B) = 2\bigg(\frac{(2A-1)}{(2A-1-2Bz(x))}-\frac{[(2A-1)^2-B^2]}{(2A-1-2Bz(x))^2}\bigg). 
\ee   
The solutions of the Schr\"odinger equation corresponding to this potential 
are given in the form of exceptional Jacobi Polynomials 
$(\hat{P}^{(\alpha,\beta)}_{n+1}(g(x)))$ as \cite{que}
\be\label{gvfn}
\tilde{\Psi}^{(1)}_{ext,n}(x) = N^{(1)}_{ext,n}(\alpha ,\beta)\frac{(1-z(x))^{\frac{(A-B)}{2}}(1+z(x))^{\frac{(A+B)}{2}}}{P^{(-\alpha-1,\beta-1)}_{1}(z(x))}\hat{P}^{(\alpha,\beta)}(z(x)) 
\ee
where the normalization constant is 
\be\label{khgc}
N^{(1)}_{ext,n}(\alpha,\beta) = \Bigg[\frac{n!(n+\alpha+1)(\alpha+\beta+2n+1)\Gamma (n+\alpha+\beta+1)}{2^{\alpha+\beta+1}(n+\alpha)(n+1+\beta)\Gamma (n+\alpha+1)\Gamma (n+\beta)}\Bigg]^{\frac{1}{2}}.
\ee
Here $P^{(-\alpha-1,\beta-1)}_{1}(z)$ is the classical Jacobi Polynomial for 
$n=1$.

The spectrum $\varepsilon^{2}=\tilde{E}^{(1)}_{ext,n}$ is however unchanged 
compared to the conventional Scarf case and is given by Eq. (\ref{vfcg}).

(b) {\bf{The $X_{m}$ Case:}}

Here, we replace $\tilde{\phi}(x)\longrightarrow \tilde{\phi}_{m,ext}(x,A,B)$ 
given by
\be\label{lkj}
\tilde{\phi}_{m,ext}(x,A,B)= \tilde{\phi}_{con}(x,A,B)+ \tilde{\phi}_{m,rat}(x,A,B),
\ee
where $\tilde{\phi}_{con}(x,A,B)$ is again given by Eq. (\ref{dsr}) while 
$\tilde{\phi}_{m,rat}(x,A,B)$ is given in terms of the Jacobi polynomials by 
\be\label{jhiy}
\tilde{\phi}_{m,rat}(x,A,B) = -\frac{(\beta-\alpha+m-1)}{2}z^{'}(x)\bigg[\frac{P^{(-\alpha-1,\beta+1)}_{m-1}(z(x))}{P^{(-\alpha-2,\beta)}_{m}(z(x))}-\frac{P^{(-\alpha,\beta)}_{m-1}(z(x))}{P^{(-\alpha-1,\beta-1)}_{m}(z(x))}\bigg].
\ee
Using Eq. (\ref{wsa}), the corresponding potential 
$\tilde{V}^{(1)}_{ext,m}(x,A,B)$ turns out to be 
\be\label{pjbk1}
\tilde{V}^{(1)}_{ext,m}(x,A,B) =\tilde{V}_{con}(x,A,B)
+\tilde{V}_{m,rat}(x,A,B)\,,
\ee
where $\tilde{V}_{con}(x,A,B)$ is again given by Eq. (\ref{kljh}) while the $m$ 
dependent rational potential is  
\ba\label{uijk1}
\tilde{V}^{(1)}_{m,rat}(x,A,B)&=&(2B-m-1)[2A-1+(-2B+1)z(x)]\bigg(\frac{P^{(-\alpha,\beta)}_{m-1}(z(x))}{P^{(-\alpha-1,\beta-1)}_{m}(z(x))}\bigg)\nonumber\\
&+&\frac{(-2B-m+1)^2}{2}(z'(x))^{2}\bigg(\frac{P^{(-\alpha,\beta)}_{m-1}(z(x))}{P^{(-\alpha-1,\beta-1)}_{m}(z(x))}\bigg)^2\nonumber\\
&-&2m(-2B-m-1); \quad -\pi/2<x<\pi/2,\quad 0<B<A-1.
\ea
The eigen functions $\tilde{\Psi} ^{(1)}_{ext,n,m}(x,A,B)$ of the Schr\"odinger
equation with this potential 
turn out to be
\be\label{mrfdt} 
\tilde{\Psi} ^{(1)}_{ext,n,m}(x,A,B)= N^{(1)}_{ext,n,m}(\alpha ,\beta )\frac{(1-z(x))^{\frac{(A-B)}{2}}(1+z(x))^{\frac{(A+B)}{2}}}{P^{(-\alpha-1,\beta-1)}_{m}(z(x))}
\hat{P}^{(\alpha,\beta)}_{n+m}(z(x))  
\ee
where
\be\label{khgc1}
N^{(1)}_{ext,n,m}(\alpha,\beta)=\Bigg[\frac{n!(n+\alpha+1)^{2}
(\alpha+\beta+2n+1)\Gamma (n+\alpha+\beta+1)}{2^{\alpha+\beta+1}(n+\alpha-m+1)
(n+m+\beta)\Gamma (n+\alpha+2)\Gamma (n+\beta)}\Bigg]^{\frac{1}{2}}\,,
\ee
while the $X_m$ exceptional Jacobi polynomials satisfy 
\ba
\hat{P}^{(\alpha,\beta)}_{n+m}(z)=(-1)^m\bigg[\frac{1+\alpha+\beta+n}{2(1+\alpha+n)}(g-1)P^{(-\alpha-1,\beta-1)}_{m}(g)P^{(\alpha+2,\beta)}_{n-1}(g)\nonumber \\+\frac{1+\alpha-m}{\alpha+1+n}P^{(-2-\alpha,\beta)}_{m}(g)P^{(\alpha+1,\beta-1)}_{n}(g)\bigg];\quad n,m\geq 0.
\ea 
The energy spectrum is again same as that of the conventional or $X_{1}$ case
and is given by Eq. (\ref{vfcg}).
 
The plots of the Dirac scalar 
potentials $\tilde{\phi}_{ext,m}(x,A,B)$ and   
the corresponding normalized ground state eigen functions 
$\tilde{\Psi}^{(1)}_{ext,0,m}(x,A,B)$ are shown for $m = 0, 1, 2$ in 
figs. $2(a)$ and $2(b)$ respectively in case $A=3$ and $B=1$.

\includegraphics[scale=1.2]{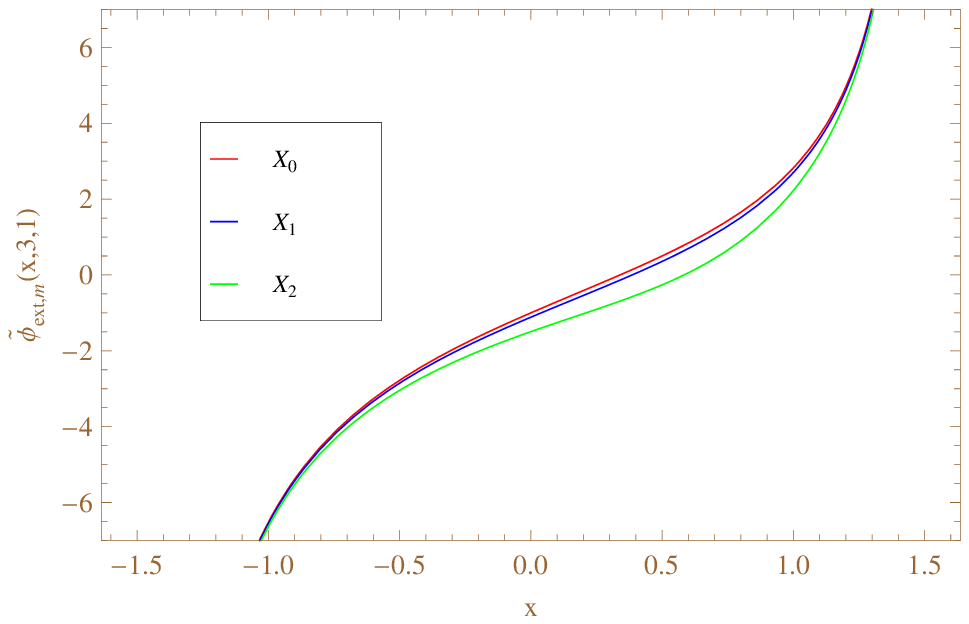}\\ 
{\bf Fig.2}: {(a) {\it Rationally extended Dirac scalar potentials for $m=0,1$ and $2$.}\\\
\includegraphics[scale=1.2]{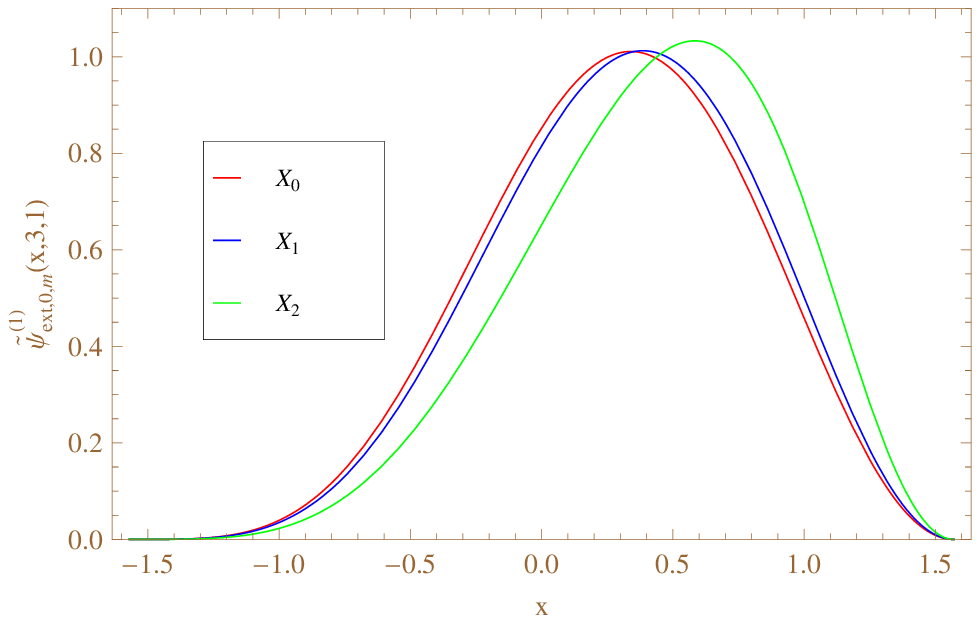} \\
{\bf Fig.2}: {(b) {\it Normalized ground state wave functions for $m=0,1$ and $2$.}\\\ 

\subsection{Hyperbolic P\"oschl-Teller case}

\subsubsection{The Conventional Case}

In this case we define 
\be\label{boki}
\tilde{\phi}(r)\longrightarrow \tilde{\phi}_{con}(r,A,B) = A\coth{r} -B\cosech{r}, \quad 0\leq r\leq \infty 
\ee
with $B>A+1>1$ which gives rise to the conventional hyperbolic 
P\"oschl-Teller potential \cite{khare}
\be\label{xszc}
\tilde{V}^{(1)}_{con}(r)=[(A+1)A+B^{2}]\cosech^{2}r
-B(2A+1)\cosech{r}\coth{r}+A^{2}\,.
\ee
The corresponding eigen functions of the Schr\"dinger equation are
\be\label{mjhg} 
\tilde{\Psi} ^{(1)}_{con}(r,A,B) = N^{(1)}_{con,n}(\alpha,\beta)\big(z(r)-1\big)^{\frac{(B-A)}{2}}\big(z(r)+1\big)^{-\frac{(B+A)}{2}}P^{(\alpha,\beta)}_{n}(z(r)).  
\ee
where $\alpha = -A+B-\frac{1}{2}$, $\beta=-A-B-\frac{1}{2}$, $z(r)=\cosh{r}$ 
and the normalization constant is
\be\label{mjhbv}
N^{(\alpha,\beta)}_{con}=\bigg[\frac{n!(-\alpha-\beta-2n-1)(n+\alpha+1)(\alpha+n+1)\Gamma(-\beta-n)}{2^{\alpha+\beta+1}(\alpha+1)^2\Gamma(\alpha+n+1)\Gamma(-\alpha-\beta-n)}\bigg]^{1/2}.
\ee
The energy eigenvalue spectrum turns out to be 
\be\label{hgb}
\varepsilon^{2}=\tilde{E}^{(1)}_{con,n} = A^2-(A-n)^2\ , \quad n = 0,1,2,...,n_{max}<A.
\ee

\subsubsection{The Extended Case}

(a) {\bf{The $X_{1}$ Case:}} 

In this case, the Dirac scalar potential is defined as
\be\label{uygf}
\tilde{\phi}_{ext}(r,A,B)= \tilde{\phi}_{con}(r,A,B) + \tilde{\phi}_{rat}(r,A,B)
\ee
where $\tilde{\phi}_{con}(r,A,B)$ is given by Eq. (\ref{boki}) while 
\be\label{wugpt}
\tilde{\phi}_{rat}(r,A,B) = 2Bz^{'}(r)\bigg[\frac{1}{2Bz(r)-2A-1}-\frac{1}{2Bz(r)-2A+1}\bigg].
\ee
Using Eq. (\ref{wsa}), this leads to the rationally extended hyperbolic 
P\"oschl-Teller potential\cite{que,scatt}
\be\label{ouk}
\tilde{V}^{(1)}_{ext}(r,A,B)=\tilde{V}^{(1)}_{con}(r,A,B)
+\tilde{V}^{(1)}_{rat}(r,A,B)
\ee
where $\tilde{V}^{(1)}_{con}(r,A,B)$ is given by Eq. (\ref{xszc}) while
\be\label{oihy}
\tilde{V}^{(1)}_{rat}(r,A,B)=2\bigg[\frac{(2A+1)}{(2Bz(r)-2A-1)}
-\frac{(4B^2-(2A+1)^2)}{(2Bz(r)-2A-1)^2}\bigg]+A^2\,.
\ee 
The corresponding eigen functions of the Schr\"odinger equation turn out to be 
\cite{que,scatt}
\be\label{mjhg1} 
\tilde{\Psi} ^{(1)}_{ext,n}(r,A,B)=N^{(1)}_{ext,n}(\alpha,\beta)\frac{(z(r)-1)^{\frac{(B-A)}{2}}(z(r)+1)^{-\frac{(B+A)}{2}}}{P^{(-\alpha-1,\beta-1)}_{1}(z(r))}\hat{P}^{(\alpha,\beta)}_{n+1}(z(r)),  
\ee
where the normalization constant is
\be\label{mgx}
N^{(1)}_{ext,n}(\alpha,\beta)=\bigg[\frac{n!(-\alpha-\beta-2n-1)(\alpha+n+1)\Gamma(-\beta-n+1)}{2^{\alpha+\beta+1}(-\beta-n-1)(\alpha)^2
\Gamma(\alpha+n)\Gamma(-\alpha-\beta-n)}\bigg]^{1/2}\,.
\ee
The energy spectrum is same as that of the conventional case and given by 
Eq. (\ref{hgb}).

(b) {\bf{The $X_{m}$ Case:}} 

In this case the Dirac scalar potential (for any arbitrary $m$) is given by
\be\label{jbnh}
\tilde{\phi}_{ext,m}(r,A,B) = \tilde{\phi}_{con}(r,A,B)
+ \tilde{\phi}_{m,rat}(r,A,B)\,,
\ee
where $\tilde{\phi}_{con}(r,A,B)$ is again given by Eq. (\ref{boki}) while
\be\label{wrgpt}
\tilde{\phi}_{m,rat}(r,A,B) = -\frac{(\beta -\alpha + m-1)}{2}z'(r)\bigg[\frac{P^{(-\alpha -1,\beta +1)}_{m-1}(z(r))}{P^{(-\alpha -2,\beta )}_{m}(z(r))}-\frac{P^{(-\alpha ,\beta )}_{m-1}(z(r))}{P^{(-\alpha -1,\beta -1)}_{m}(z(r))}\bigg].
\ee
Using Eq. (\ref{wsa}) this leads to the potential which is now $m$-dependent 
\cite{os,scatt} and is given as
\be\label{ygth}
\tilde{V}^{(1)}_{ext,m}(r,A,B)=\tilde{V}^{(1)}_{con}(r,A,B)
+\tilde{V}^{(1)}_{rat,m}(r,A,B)\,,
\ee
where $\tilde{V}^{(1)}_{con}(r,A,B)$ is again given by Eq. (\ref{xszc}) while
\ba\label{gffc}
\tilde{V}^{(1)}_{rat,m}(r,A,B)&-&(2B-m+1)[2A+1-(2B+1)z(r)]\bigg(\frac{P^{(-\alpha,\beta)}_{m-1}(z(r))}{P^{(-\alpha-1,\beta-1)}_{m}(z(r))}\bigg)\nonumber\\
&+&\frac{(2B-m+1)^2}{2}(z'(r))^{2} \bigg(\frac{P^{(-\alpha,\beta)}_{m-1}(z(r))}{P^{(-\alpha-1,\beta-1)}_{m}(z(r))}\bigg)^2\nonumber\\
&+&2m(-2B-m-1); \quad 0\leq x\leq \infty, \quad B>A+1>1.
\ea 
The corresponding eigen functions of the Schr\"odinger equation turn out to be 
\be\label{mjhg2} 
\tilde{\Psi} ^{(1)}_{ext,n,m}(r,A,B)=N^{(1)}_{ext,n,m}(\alpha,\beta)\frac{(z-1)^{\frac{(B-A)}{2}}(z+1)^{-\frac{(B+A)}{2}}}{P^{(-\alpha-1,\beta-1)}_{m}(z(r))}\hat{P}^{(\alpha,\beta)}_{n+m}(z(r)), 
\ee

where the  normalization constant is given by 
\be\label{mjhbv1}
N^{(1)}_{ext,n,m}(\alpha,\beta)=N^{(1)}_{ext,n}(\alpha ,\beta )\bigg(\frac{(n+\alpha -m+1)}{(\alpha +n)}\bigg)^{\frac{1}{2}}.
\ee
The energy eigenvalue spectrum is again same as in the conventional case and
is given by Eq. (\ref{hgb}).

The plots of the Dirac scalar 
potentials $\tilde{\phi}_{ext,m}(r,A,B)$ and   
the corresponding normalized ground state eigen functions 
$\tilde{\Psi}^{(1)}_{ext,0,m}(r,A,B)$ are shown for $m = 0, 1, 2$ in 
figs. $3(a)$ and $3(b)$ respectively in case $A=1$ and $B=3$.

\includegraphics[scale=1.2]{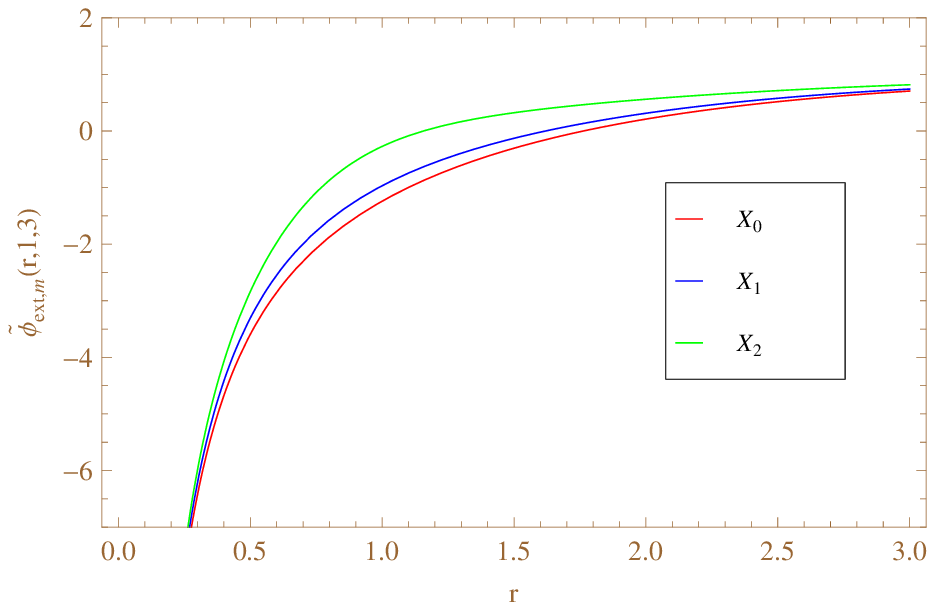}\\ 
{\bf Fig.3}: {(a) {\it Rationally extended Dirac scalar potentials for $m = 0,1$ and $2$.}\\\
\includegraphics[scale=1.2]{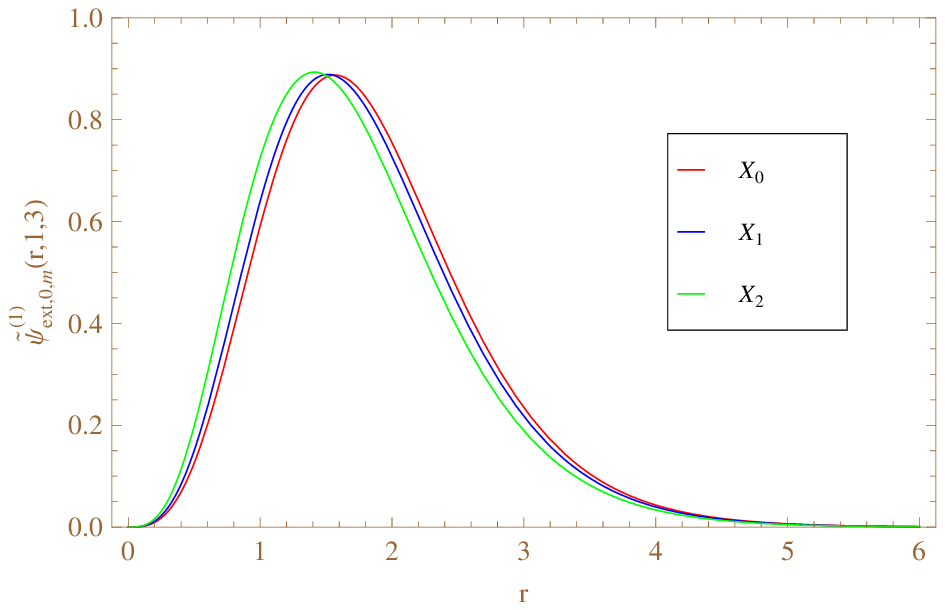} \\
{\bf Fig.3}: {(b) {\it Normalized ground-state wave functions for $m = 0, 1$ and $2$.}\\\ 

\section{Parametric symmetry and new forms of Dirac Scalar Potentials}

Recently, the role of the parametric symmetry has been discusses in the case of
Schr\"odinger equation \cite{para2,para1}. It is then worthwhile to discuss the
role of the parametric symmetry in the context of the Dirac equation. 
It turns out that out of three examples discussed in this paper,  this symmetry
only exist in the two cases, i.e. the trigonometric scarf and the hyperbolic 
P\"oschl-Teller 
potentials which we discuss one by one. We shall see that in the
conventional cases, this symmetry generates new Dirac scalar potentials
keeping the corresponding Schr\"odinger potential $\tilde{V}^{(1)}_{con}$ unchanged but having
a new partner $\tilde{V}^{(2)}_{con}$. On the other hand, in the extended cases, one finds that
the corresponding $\tilde{V}^{(1)}_{ext}$ and $\tilde{V}^{(2)}_{ext}$ are both modified.

\subsection{Trigonometric Scarf Case}

\subsubsection{The Conventional Case}

If we replace the parameters $B \longleftrightarrow A-\frac{1}{2}$ in the conventional scalar potential 
$\tilde{\phi}_{con}(x,A,B)$ as given in \cite{para2}, we get a new form of 
$\tilde{\phi}^{(p)}_{con}(x,A,B)$ generated due to this parametric 
transformation i.e., 
\ba\label{ihgy}
\tilde{\phi}(x)\rightarrow \tilde{\phi}^{(p)}_{con}(x,A,B)&=&\tilde{\phi}_{con}(x,A\rightarrow B+\frac{1}{2},B\rightarrow A-\frac{1}{2})\nonumber\\
&=&\bigg(B+\frac{1}{2}\bigg)\tan{x}-\bigg(A-\frac{1}{2}\bigg)\sec{x};\quad B>A-1>0\nonumber\\
\ea 
which is different from the $\tilde{\phi}_{con}(x,A,B)$ as given by Eq. 
(\ref{dsr}). Remarkably, this scalar potential leads to
the same potential $\tilde{V}^{(1,p)}_{con}(x,A,B) 
=\tilde{V}^{(1)}_{con}(x,A,B)$  as given by Eq. (\ref{kljh}) but different 
$\tilde{V}^{(2,p)}_{con}(x)=\tilde{V}_{con}(x,A,B\rightarrow B+1)$. 
In other words, $\tilde{V}^{(1)}_{con}(x,A,B)$ has two different SUSY partners.
Thus we have another set of $\tilde{\phi}(i.e,\tilde{\phi}^{(p)}_{con}(x,A,B))$
leading to different eigenvalues and eigen functions,  i.e,
\be\label{wfscx1}
\tilde{\Psi}^{(1,p)}_{con,n}(x,A,B)=\tilde{\Psi}^{(1)}_{con,n}(x,A\rightarrow B+\frac{1}{2},B\rightarrow A-\frac{1}{2});\quad B>A-1>0
\ee

and
\be\label{wfv1}
\tilde{\Psi}^{(2,p)}_{n}(x,A,B)=\tilde{\Psi}^{(1,p)}_{con,n}(x,A,B\rightarrow B+1),
\ee
with the energy eigenvalues 
\be\label{etf}
\varepsilon ^{2}=\tilde{E}_{n}^{(1,p)}=(B+n+\frac{1}{2})^{2}, \quad n=0,1,2,...
\ee

\subsubsection{The extended Case}

(a) {\bf{The $X_{1}$ case:}} 

In the extended case as $B\longleftrightarrow A-\frac{1}{2}$,
we have  
\ba\label{sna}
\tilde{\phi}(x)&\longrightarrow & \tilde{\phi}_{ext}^{(p)}(x,A,B)=\tilde{\phi}_{ext}(x,A\rightarrow B+\frac{1}{2},B\rightarrow A-\frac{1}{2})\nonumber\\
&=& \tilde{\phi}_{con}^{(p)}(x,A,B)+\tilde{\phi}_{rat}^{(p)}(x,A,B),
\ea
where $\tilde{\phi}_{con}^{(p)}(x,A,B)$ is as given by Eq. (\ref{ihgy}) while
 


\be\label{lkvk}
\tilde{\phi}_{rat}^{(p)}(x,A\rightarrow B+\frac{1}{2},B\rightarrow A
-\frac{1}{2})= 2\bigg(A-\frac{1}{2}\bigg)z'(x)\bigg[\frac{1}{2B+2-(2A-1)z(x)}
-\frac{1}{2B-(2A-1)z(x)}\bigg]\,.
\ee


Note that this scalar extended Dirac potential Eq. (\ref{sna}) is different 
from (\ref{ppj}) and unlike the conventional case it leads to both  
$\tilde{V}_{ext}^{(1,p)}(x,A,B)$ and $\tilde{V}_{ext}^{(2,p)}(x,A,B)$ being 
different and are given as 
\be\label{uyhg}
\tilde{V}^{(1,p)}_{ext}(x,A,B)=\tilde{V}^{(1)}_{ext}(x,A\rightarrow B+\frac{1}{2},B\rightarrow A-\frac{1}{2})
\ee
and
\be\label{uyhg1}
\tilde{V}^{(2,p)}_{ext}(x,A,B)=\tilde{V}^{(1,p)}_{ext}(x,A,B\rightarrow B+1)
\ee
The corresponding eigen functions of the Schr\"odinger equation can be written as
\ba\label{wfscx1}
\tilde{\Psi}^{(1,p)}_{ext,n}(x,A,B)&=&\tilde{\Psi}^{(1)}_{ext,n}(x, A\rightarrow B+\frac{1}{2}, B\rightarrow A-\frac{1}{2});\quad B>A-1>0\nonumber\\
&=&N^{(1,p)}_{ext}(\gamma ,\delta )\frac{(1-z(x))^{\frac{(B-A+1)}{2}}(1+z(x))^{\frac{(A+B)}{2}}}{(2B-2(A-\frac{1}{2})z(x))}\hat{P}^{(\gamma ,\delta)}_{n+1}(z(x))
\ea
and
\be\label{wfscjh1}
\tilde{\Psi}^{(2,p)}_{ext,n}(x,A,B)
=\tilde{\Psi}^{(1,p)}_{ext,n}(x,A,B\rightarrow B+1)\,,
\ee
where $\gamma = B-A+\frac{1}{2}$ and $\delta = A+B-\frac{1}{2}$.


The corresponding energy eigenvalues are however unchanged from the 
conventional case and are again given by Eq. (\ref{etf}). 

(b) {\bf{The $X_{m}$ Case:}}

For the more general $X_m$ case, in the extended case, as 
$B\longleftrightarrow A-\frac{1}{2}$, the Dirac scalar potential is
\be\label{evp}
\tilde{\phi}^{(p)}_{ext,m}(x,A,B)=\tilde{\phi}^{(p)}_{con}(x,A,B)+\tilde{\phi}^{(p)}_{rat,m}(x,A,B)
\ee
where $\tilde{\phi}^{(p)}_{con}(x,A,B)$ is again given by Eq. (\ref{ihgy}) 
while


 
\ba\label{ygf}
\tilde{\phi}^{(p)}_{rat,m}(x,A,B)&=&\tilde{\phi}_{rat,m}(x,A\rightarrow B+\frac{1}{2},B\rightarrow A-\frac{1}{2} )\nonumber\\
&=&-\bigg(\frac{2A+m-2}{2}\bigg)z'(x)\bigg[\frac{P^{(-\alpha  -1,\beta  +1)}_{m-1}(z(x))}{P^{(-\alpha -2,\beta  )}_{m}(z(x))}
-\frac{P^{(-\alpha ,\beta )}_{m-1}(z(x))}{P^{(-\alpha -1,\beta  -1)}_{m}(z(x))}\bigg].\nonumber\\
\ea
This scalar extended potential leads to both  
$\tilde{V}_{ext}^{(1,p)}(x,A,B)$ and $\tilde{V}_{ext}^{(2,p)}(x,A,B)$ being 
different.



The corresponding eigen functions of the Schr\"odinger equation can be written 
as 
\be\label{rdgf}
\tilde{\Psi}^{(1,p)}_{ext,m,n}(x,A,B)=\tilde{\Psi}^{(1)}_{ext,m,n}(x,A\rightarrow B+\frac{1}{2},B\rightarrow A-\frac{1}{2});\quad B>A-1>0
\ee
and
\be\label{wfscjh2}
\tilde{\Psi}^{(2,p)}_{ext,m,n}(x,A,B)=\tilde{\Psi}^{(1,p)}_{ext,m,n}(x,A,B\rightarrow B+1)
.\ee



The energy eigenvalues are however unchanged and are again given by 
Eq. (\ref{etf}). 

The plots of the scalar potential $\phi^((p))_{ext,m}(x,A,B)$ and the 
normalized ground state eigen function $\tilde{\Psi}^{(1,p)}_{ext,0,m}(x,A,B)$ 
for  $A=\frac{3}{2}, B=\frac{5}{2}$ and different values of $m (=0,1$ and $2)$ 
corresponding to the conventional, $X_{1}$ and $X_{2}$ respectively  
are given in Figs. $4(a)$ and $4(b)$ respectively.\\

\includegraphics[scale=1.2]{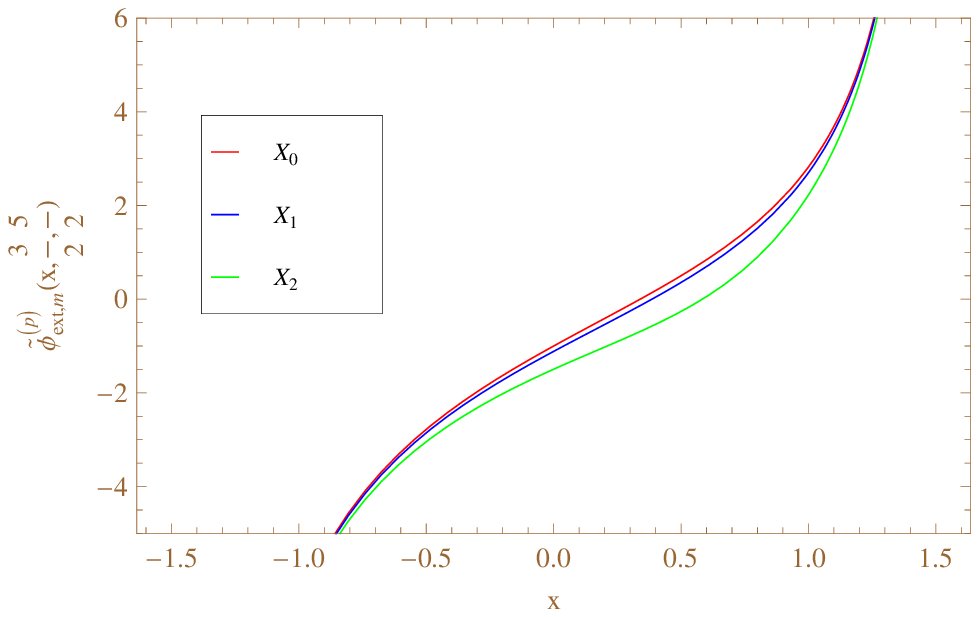}\\ 
{\bf Fig.4}: {(a) {\it Rationally extended parametric Dirac scalar potentials for $m = 0,1$ and $2$.}\\\
\includegraphics[scale=1.2]{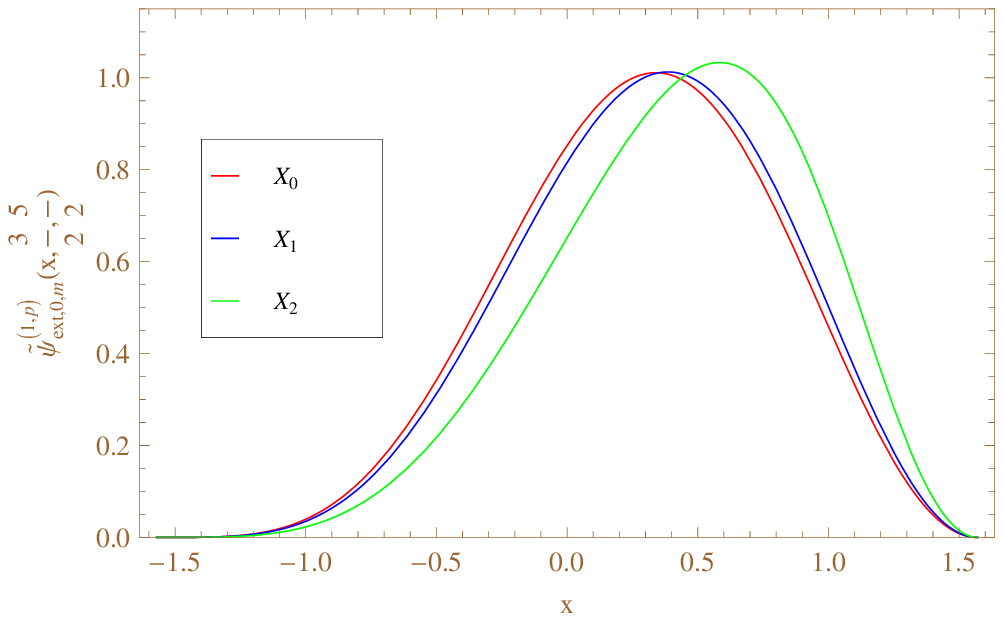} \\
{\bf Fig.4}: {(b) {\it Normalized ground-state wave functions for $m = 0, 1$ and $2$.}\\\ 

\subsection{Hyperbolic P\"oschl-Teller Case}

\subsubsection{The Conventional Case}

Similar to the  trigonometric Scarf case, in this case under the parametric 
transformation $B\leftrightarrow A+\frac{1}{2}$ 
the scalar potential $\tilde{\phi}_{con}(r,A,B)$ as given by Eq. 
(\ref{boki}) becomes
\ba\label{wtfb}
\tilde{\phi}_{con}^{(p)}(r,A,B)&=&\tilde{\phi}_{con}(r,A\rightarrow B-\frac{1}{2},B\rightarrow A+\frac{1}{2})\nonumber\\
&=&\big(B-\frac{1}{2}\big)\coth{r} -\big(A+\frac{1}{2}\big)\cosech{r};\quad 0< r< \infty.
\ea
Remarkably, under this transformation, the potential 
$\tilde{V}^{(1,p)}_{con}(r,A,B) 
=\tilde{V}^{(1)}_{con}(r,A,B)$ remains the same as given by Eq. (\ref{xszc}), 
however the partner potential $\tilde{V}^{(2,p)}_{con}(r,A,B)$ gets changed i.e,
\be\label{ptyg}
\tilde{V}^{(1,p)}_{con}(r,A,B)=\tilde{V}^{(1)}_{con}(r,A\rightarrow B-\frac{1}{2},B\rightarrow A+\frac{1}{2})=\tilde{V}^{(1)}_{con}(r,A,B)
\ee
and
\be\label{cvf}
\tilde{V}^{(2,p)}_{con}(r,A,B)=\tilde{V}^{(1,p)}_{con}(r,A,B\longrightarrow B-1)
\ee
In other words, $\tilde{V}^{(1)}_{con}(x,A,B)$ has two different SUSY partners.
The eigen functions $\tilde{\Psi}^{(1,p)}_{con,n}(r,A,B)$ and 
$\tilde{\Psi}^{(2,p)}_{con,n}(r,A,B)$ are different from the conventional 
case and related to them by
\be\label{lrau}
\tilde{\Psi}^{(1,p)}_{con,n}(r,A,B)=\tilde{\Psi}^{(1)}_{con,n}(r,A\rightarrow B-\frac{1}{2},B\rightarrow A+\frac{1}{2}),\quad A>-\frac{1}{2}, \quad B>0
\ee
and
\be\label{lqws}
\tilde{\Psi}^{(2,p)}_{con,n}(r,A,B)=\tilde{\Psi}^{(1,p)}_{con,n}(r,A,B\rightarrow B-1)
\ee
The corresponding energy eigenvalue are
\be\label{phug}
\tilde{E}_{n}^{(1,p)}=\varepsilon ^{2} = -(B-n-\frac{1}{2})^{2}, \quad n = 0,1,2,...,n_{max}<B-\frac{1}{2}.
\ee

\subsubsection{The extended Case}

(a) {\bf{The $X_{1}$ Case:}}

In this case, under the transformation $B\leftrightarrow A+\frac{1}{2}$ the 
extended scalar potential is given by 
\ba\label{rwas}
\tilde{\phi}^{(p)}_{ext}(r,A,B)&=&\tilde{\phi}_{ext}(r,A\rightarrow B-\frac{1}{2},B\rightarrow A+\frac{1}{2})\nonumber\\
&=&\tilde{\phi}^{(p)}_{con}(r,A,B)+\tilde{\phi}^{(p)}_{rat}(r,A,B)
\ea
where $\tilde{\phi}^{(p)}_{con}(r,A,B)$ is as given by Eq. (\ref{boki}) while 
\ba\label{yftg}
\tilde{\phi}^{(p)}_{rat}(r,A,B)&=&\tilde{\phi}_{rat}(r,A\rightarrow B-\frac{1}{2},B\rightarrow  A+\frac{1}{2})\nonumber\\
&=&2\bigg(A+\frac{1}{2}\bigg)z'(x)\bigg[\frac{1}{2(A+\frac{1}{2})z(r)-2B}-\frac{1}{2(A+\frac{1}{2})z(r)-2B+2}\bigg].\nonumber\\
\ea
The extended potentials under this transformation are completely different and
are given by
\be\label{lkv}
\tilde{V}^{(1,p)}_{ext}(r,A,B)=\tilde{V}^{(1)}_{ext}(r,A\rightarrow B-\frac{1}{2},B\rightarrow A+\frac{1}{2})
\ee
and
\be\label{mjlh}
\tilde{V}^{(2,p)}_{ext}(r,A,B)=\tilde{V}^{(1,p)}_{ext}(r,A,B\rightarrow  B-1).
\ee
The associated eigen functions of the Schr\"odinger equation are 
\be\label{kgfh}
\tilde{\Psi}^{(1,p)}_{ext,n}(r,A,B)=\tilde{\Psi}^{(1)}_{ext,n}(r,A\rightarrow B-\frac{1}{2},B\rightarrow A+\frac{1}{2})
\ee
and
\be\label{kgz}
\tilde{\Psi}^{(2,p)}_{ext,n}(r,A,B)
=\tilde{\Psi}^{(1,p)}_{ext,n}(r,A,B\rightarrow B-1)\,.
\ee
The energy eigenvalues though are unchanged and are given by Eq. (\ref{phug}).\\
(b) {\bf{The $X_{m}$ Case:}}\\

For the $X_{m}$-case, we define
\be
\tilde{\phi}_{m,ext}^{(p)}(r,A,B)= \tilde{\phi}_{con}^{(p)}(r,A,B)+ \tilde{\phi}_{m,rat}^{(p)}(r,A,B)
\ee
where $\tilde{\phi}_{con}^{(p)}(r,A,B)$ is again given by Eq. (\ref{boki}) 
while
\be\label{wrgpt1}
\tilde{\phi}_{m,rat}^{(p)}(r,A,B) =\frac{[2(A+1)-m]}{2}z'(r)\bigg[\frac{P^{(-\eta  -1,\zeta  +1)}_{m-1}(z(r))}{P^{(-\eta  -2,\zeta  )}_{m}(z(r))}-\frac{P^{(-\eta  ,\zeta  )}_{m-1}(z(r))}{P^{(-\eta  -1,\zeta  -1)}_{m}(z(r))}\bigg],
\ee
here $\eta =A-B+\frac{1}{2}$ and $\zeta =-A-B-\frac{1}{2}$. 
The extended potentials under this transformation are again completely 
different. The associated eigenfunctions of the Schr\"odinger equation are 
\be\label{kgfh1}
\tilde{\Psi}^{(1,p)}_{ext,m,n}(r,A,B)=\tilde{\Psi}^{(1)}_{ext,m,n}(r,A\rightarrow B-\frac{1}{2},B\rightarrow A+\frac{1}{2});\quad A+1>B>0
\ee
and
\be\label{kgz1}
\tilde{\Psi}^{(2,p)}_{ext,m,n}(r,A,B)=\tilde{\Psi}^{(1,p)}_{ext,m,n}(r,A, B\rightarrow B-1).
\ee
 The energy eigenvalues are remain unchanged and given by Eq. (\ref{phug}). The plots of the Dirac scalar potential $\phi^((p))_{ext,m}(r,A,B)$ and the 
 normalized ground state eigen functions $\tilde{\Psi}^{(1,p)}_{ext,0,m}(r,A,B)$ 
 for the parameters $A=\frac{5}{2}, B=\frac{3}{2}$ and  $m=0,1, 2$  
are shown in Figs. $5(a)$ and $5(b)$ respectively.\\

\includegraphics[scale=1.2]{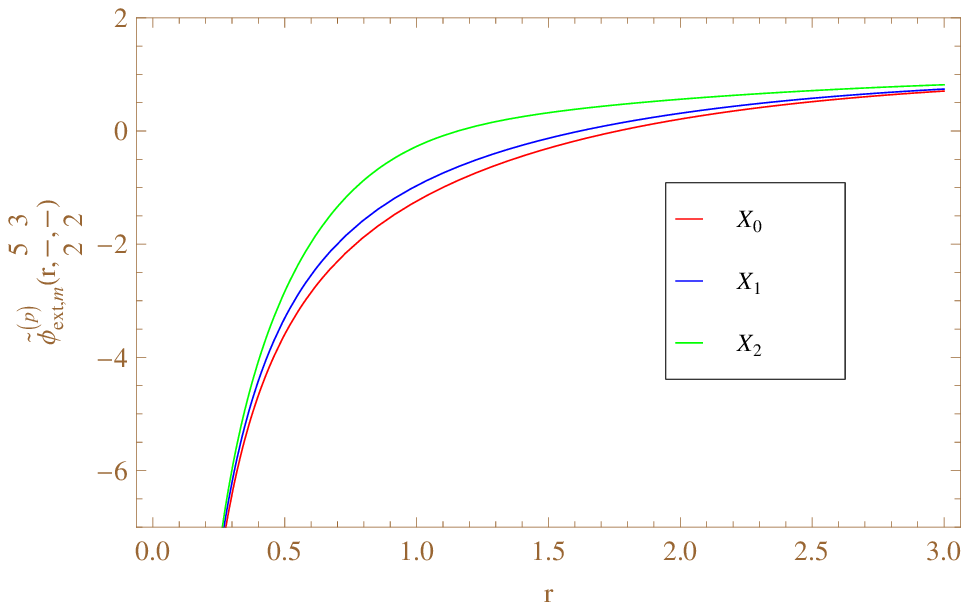}\\ 
{\bf Fig.5}: {(a) {\it Rationally extended parametric Dirac scalar potentials for $m = 0,1$ and $2$.}\\\
\includegraphics[scale=1.2]{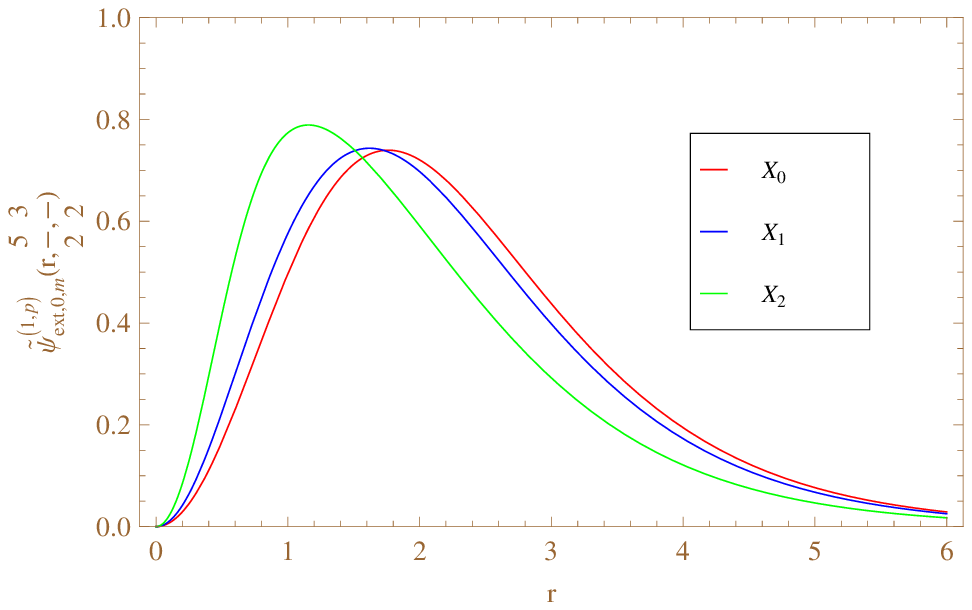} \\
{\bf Fig.5}: {(b) {\it Normalized ground-state wave functions for $m = 0, 1$ and $2$.}\\\ 

\section{Summary and Possible Open Problems}

In this paper we have obtained exact solutions of the $1+1$-dimensional Dirac 
equation for three different extended scalar potentials, i.e. radial 
oscillator, trigonometric Scarf and hyperbolic Poschl-Teller potentials 
in terms of exceptional orthogonal polynomials by connecting them to the 
corresponding Schr\"odinger problems. 
Further, using the idea of the parametric symmetry in the case of the 
trigonometric Scarf and the hyperbolic P\"oschl-Teller Dirac scalar potentials we 
have generated a new class of conventional as well as rational scalar 
potentials and have obtained their 
exact solutions in terms of conventional as well as exceptional orthogonal
polynomials.

This paper raises few obvious questions. For example, are there other exactly 
solvable Dirac scalar potentials whose solutions are also in terms of the 
exceptional orthogonal polynomials. Secondly, are there other Dirac scalar 
potentials admitting parametric symmetry and if yes can one obtain the
solutions of the newly generated Dirac scalar problem?  

{\bf Acknowledgments}\\
AK is grateful to Indian National Science Academy (INSA) for 
awarding INSA Honorary Scientist position at Savitribai Phule Pune University. BPM acknowledges the research grant for faculty under IoE scheme (Number 6031) of  
Banaras Hindu University Varanasi.

\end{document}